# Base rate neglect in computer science education

**Koby Mike**

Faculty of Education in Science and Technology, Technion – Israel Institute of Technology, Haifa, Israel.

kobymike@gmail.com

**Orit Hazzan**

Faculty of Education in Science and Technology, Technion – Israel Institute of Technology, Haifa, Israel.

oritha@technion.ac.il

## ABSTRACT

Machine learning (ML) algorithms are gaining increased importance in many academic and industrial applications, and such algorithms are, accordingly, becoming common components in computer science curricula. Learning ML is challenging not only due to its complex mathematical and algorithmic aspects, but also due to a) the complexity of using correctly these algorithms in the context of real-life situations and b) the understanding of related social and ethical issues.

Cognitive biases are phenomena of the human brain that may cause erroneous perceptions and irrational decision-making processes. As such, they have been researched thoroughly in the context of cognitive psychology and decision making; they do, however, have important implications for computer science education as well. One well-known cognitive bias, first described by Kahneman and Tversky, is the base rate neglect bias, according to which humans fail to consider the base rate of the underlying phenomena when evaluating conditional probabilities.

In this paper, we explore the expression of the base rate neglect bias in ML education. Specifically, we show that about one third of students in an Introduction to ML course, from varied backgrounds (computer science students and teachers, data science, engineering, social science and digital humanities), fail to correctly evaluate ML algorithm performance due to the base rate neglect bias. This failure rate should alert educators and promote the development of new pedagogical methods for teaching ML algorithm performance.

CCS CONCEPTS • CCS → Mathematics of computing → Probability and statistics → Probabilistic inference problems → Bayesian computation • CCS → Social and professional topics → Professional topics → Computing education

**Additional Keywords and Phrases:** Artificial intelligence, Base-rate neglect, Cognitive Psychology, Cognitive bias, Machine learning.

# 1 INTRODUCTION

Data science is a new field of research that focuses on turning raw data into understanding, insight, knowledge, and value [1], [2]. It is an interdisciplinary field that integrates knowledge and methods from computer science, mathematics, and statistics, as well as from various application domains that gives meaning to the data. The radical growth in recent years in the availability of both data and the computational resources required to process them has led to a corresponding increase in the demand for data scientists. As a result, new data science education programs are being launched at a growing rate, many of which are offered to undergraduate students [3]. As data science is interdisciplinary, data science courses and programs are opened to a variety of learners, including undergraduates majoring in computer science or data science, as well as students from other domains minoring in computer science or data science, or graduate students.

Machine learning (ML) is an important component in data science. ML algorithms are using to model the data, find new connections between variables, and predict the future. ML is one of the effective methods for modeling huge and complex data and is considered to be a field of research located in the intersection of statistics and computer science. Unlike traditional parametric modeling, which presumes an underlying statistical behavior of the researched phenomena and aims to fit the best parameters of the sought after model, ML models learn from experience and can learn complex data patterns directly from raw data [4].

One of the main concerns in using ML algorithms is ML explainability and interpretability. Several definitions have been proposed for these terms and no consensus has been reached yet [5]. Both terms concern with the humans' ability to understand the predictions of ML algorithms and the reasons for these predictions [6]. Developing both interpretable models and explanation methods is currently one of the main efforts of the ML research community [7], [8] as interpretability methods are required to help users build trust in and understand the capabilities of ML models. Rudin propose that new ML techniques that will be inherently interpretable should be developed [9]. This perspective implies that there are two sides to the explainability and interpretability problem: humans and machines [10]. However, while there is an ongoing effort to improve the machines, it is the role of educators to improve human understanding on ML algorithms.

Cognitive biases are phenomena of the human brain that may cause erroneous perceptions and irrational decision-making processes [11]. In this paper, we show that difficulties in understanding the outcomes of ML algorithms, i.e., interpret its predictions and performance, may stem not only from the limitations of algorithm explainability and interpretability, but also from human cognitive biases [12].

ML is a complex topic for students [13], [14]. ML requires understanding of the algorithms themselves and the underlying mathematics, as well as a broader view of the algorithms in the context of the domain of data [15]. Such understanding requires knowledge about the role of ML as a component of the data science process, handling biases in the data, comprehension of the role of the training and tests data, the evaluation of ML methods and models in the context of the application, as well as ethics and social responsibility [16], [17].

ML explainability and interpretability is addressed mainly as a challenge for ML researchers to design interpretable ML algorithms for ML users. ML interpretability, however, is a challenge for educators as well. In our literature review we did not find research on interpretation of ML algorithms by students. Our findings reveal that students from varied backgrounds, who take introduction to ML courses, misinterpret ML algorithms outcomes. About one third of the participants in our study fail to correctly interpret the performance of ML algorithms due to the base rate neglect - neglecting the base rate probability of the underlying phenomena.

The rest of the paper is organized as follows: Section 2 presents relevant background in ML education and the base rate neglect. In Sections 3 and 4 we describe our research method and results. In Section 5 we conclude.

## 2 BACKGROUND

In this section we present the current state of research regarding the pedagogy of ML and discuss the base rate neglect.

### 2.1 Machine Learning Education

In the context of ML education, research recognizes three types of learners: majors, non-majors, and users [14], [18]. Majors are students who major in domains such as electrical engineering, computer science, or data science and have extensive mathematical background so they can learn ML algorithms with all its mathematical details. Such learning is commonly referred as learning ML as a white box [19]. Non-majors are students from other domains, such as social science or life science. Non-majors usually do not have the required mathematical background to learn ML algorithms with all its mathematical details, and thus usually learn ML algorithms as a black box. The third population are users of ML algorithms who need to understand ML as part of their professional or day to day life, i.e., physicians who use ML algorithms as a diagnostic tool or business managers who uses ML algorithms for decision making. While the goal of majors and non-majors ML curricula is to teach how the algorithms work and how to generate new ML applications, the goal of ML curricula for users is to teach the ML algorithms functionality and limitations, including the interpretation of their output and predictions.

Teaching the interpretability of ML is explicitly recommended for data science students [16] and is indeed integrated into some courses (for example [20], [21]). ML is included not only in data science curricula [16], [17], [22], it has been an advanced topic also in the curricula of computer science and electrical engineering students for at least a decade [23]. The pedagogical research on teaching ML concentrates mainly on the non-majors students (see for example [13], [14]), and interpretability is discussed mainly in the context of users. Long and Magerko [24] defines 17 competencies of AI literacy, several of which is connected to understanding the role of humans in either developing the model, choosing data for training, and examination of algorithms and bias. Suresh et al. [8] found that people trust incorrect ML recommendations for tasks that they perform correctly most of the time, even if they are knowledgeable in ML or are given information indicating the system is not confident in its prediction. Several methods have been proposed to help users interpret ML outcome. For example, Suresh et al. [25] suggest presenting a model's prediction output with other examples the user familiar with that generate the same prediction.

At the same time, Bhatt [18] explored how ML developers view and use explainability and found that the majority of developers do not concern how the end users are effected by the model, but rather they concern about the ML engineers, who use explainability to debug the model itself.

### 2.2 Machine Learning Classifiers

A classifier is an algorithm whose task is to classify objects to one of several classes. A ML classifier is an algorithm that extracts the classification rules from a labeled training data set by itself. As its name implies, a training data set contains examples that are labeled with the class they belong to. In the training phase, the machine learns the classification rule by iterating over the tagged examples, gradually updating the parameters of the classification model to minimize the classification error. For example, a classifier might be presented with 10,000 photos, of which 5,000 are tagged as cats and 5,000 are tagged as dogs. In the training phase, the classifier learns a model that can distinguish between photos of cats and dogs. After the training phase, the classifier uses this model to classify photos it has never seen before.

Classifiers might also produce false predictions, mislabeling objects of one class as objects of the other class. A common way to present classification correctness is a confusion matrix, whose rows represent the real labels and the columns represent the predicted labels. The number on the diagonal of the confusion matrix represent the numbers of the correctly classified objects; numbers that are not on the diagonal represent the number of incorrectly classified objects, since the predicted label differs from the true label. Figure 1 represents a confusion matrix of a classifier that was train to classify cats and dogs and classifies 4% of the cats as dogs (20 out of 500) and 5% of the dogs as cats (25 out of 500).

|  |  | Predicted label |  |
|---|---|---|---|
|  |  | Cats | Dogs |
| Real label | Cats | 480 | 20 |
|  | Dogs | 25 | 475 |

Figure 1: Confusion matrix

### 2.3 The Base Rate Neglect Cognitive Bias

Cognitive biases are phenomena of the human brain that may cause erroneous perceptions and irrational decision-making processes [11]. For example, consider the following question regarding the classifier presented in Figure 1: If an image is classified as an image of a cat, what is the chance that this is indeed a photo of a cat? If the proportion of cats and dogs in the population is balanced, we can tell that the classifier will classify 505 images as cats (the sum of the left column), of which 480 really are cats (the upper left cell), and so the answer is $480/505 \approx 95\%$. But what if the base rate of cats and dogs in the population is not balanced? In this case we cannot simply sum up the left column, but rather must consider the base rate of each class. Research tells, however, that humans tend to neglect the base rate when performing such calculations.

For example, Casscells et al. [26] asked 20 interns, 20 fourth-year medical students, and 20 attending physicians at four Harvard Medical School teaching hospitals the following question, referred to as the medical diagnosis problem:

> *"If a test to detect a disease whose prevalence is 1/1000 has a false positive rate of 5%, what is the chance that a person found to have a positive result actually has the disease, assuming you know nothing about the person's symptoms or signs?"*

The correct answer to this question is about 2%, based on Bayes theorem (see Equation 1). Note that in the original phrasing of the question, the conditional probability of True-Positive is not given, and is assumed to be 1, meaning that the test is positive whenever the patient has the disease.

$$p(Actually\ has\ the\ disease\ |\ Found\ positive) =$$

(Eq. 1) $$= \frac{p(Positive\ |\ Disease) \cdot p(Disease)}{p(Positive)} =$$

$$= \frac{(Positive \mid Disease) \cdot p(Disease)}{p(Positive \mid Disease) \cdot p(Disease) + p(Positive \mid No\ Disease) \cdot p(Do\ disease)} =$$

$$= \frac{1 \cdot 0.001}{1 \cdot 0.001 + 0.05 \cdot 0.999} =$$

$$= 0.0196 \approx 2\%$$

Only 18% of the participants in Casscells et al. [26] experiment gave the correct answer. Several explanations were suggested for this phenomenon [27]. The base-rate neglect fallacy, identified by Kahneman and Tversky [11], is one of these explanations, suggests that this mistake results from ignoring the base rate of the disease in the population. Another explanation was suggested by Leron and Hazzan [28], who explained this error using the dual-process theory [29].

A known method of mitigating the base-rate neglect cognitive bias is to work with frequencies (natural numbers) instead of probabilities (percentages). For example, Cosmides and Tooby [30] rephrased the medical diagnosis problem as follows:

*"One out of every 1,000 Americans has disease X. A test has been developed to detect when a person has disease X. Every time the test is given to a person who has the disease, the test comes out positive (i.e., the "true positive" rate is 100%). But sometimes the test also comes out positive when it is given to a person who is completely healthy. Specifically, out of every 1,000 people who are perfectly healthy, 50 of them test positive for the disease (i.e., the "false positive" rate is 5%). Imagine that we have assembled a random sample of 1,000 Americans.*

*They were selected by a lottery. Those who conducted the lottery had no information about the health status of any of these people. Given the information above, on average, how many people who test positive for the disease will actually have the disease? _______ out of ________ ".*

Cosmides and Tooby [30] found that 56% of the participants in their research answered the medical diagnosis problem correctly when it was formulated using frequencies. When presented in this format (with frequencies, not probabilities), it is natural to represent the information in a confusion matrix (see Figure 2). In this case, it is easy to compute the precision as 50/(50+1), without using Bayes theorem.

|  |  | Predicted label | |
| --- | --- | --- | --- |
|  |  | Detected | Not detected |
| Real label | Disease | 1 | 0 |
|  | No disease | ~50 | ~950 |

Figure 2: The medical diagnosis confusion matrix

Casscells et al. [26] and Cosmides and Tooby [30] quantified the base rate neglect bias among medical students and staff in the context of medical diagnosis. In our research, we quantify the base rate neglect bias among students taking an Introduction to ML course, in the context of ML classifiers.

## 3 METHOD

### 3.1 Research Problem and Objective

Possible effects of cognitive biases on ML interoperability have been proposed by Kliegr et al [12] but were not empirically verified. Accordingly, the objective of the current research is to empirically explore the expression of the base-rate neglect cognitive bias in the context of ML, in general, and specifically, ML education.

### 3.2 Research Question

To what extent do learners in Introduction to ML courses fail to correctly interpret ML algorithms performance due to the base rate neglect cognitive bias?

### 3.3 Data Collection

Data was collected using a questionnaire. The questionnaire included two questions regarding the base rate neglect – the lion classification question (see Table 1) and the tomato's disease question (see Table 2). The lion classification question is a ML analog of the medical diagnosis problem proposed by Casscells [26]. In this question, the students are asked to evaluate the probability of a photo to contain a lion if an ML detected a lion in this photo, that is, the true positive rate of the lion detector. The false positive rate, that is the probability of a photo that do not contains a lion to be detected as a lion photo is given as 5%. As the false negative rate (lion photos that are not detected) is given as 0, all the lion photos will be detected. The question is thus what the percentage of non-lion photos in the detected-as-lion-photos group is. This percentage depends on the ratio of lion photos in the dataset, that is, the base rate of lion photos, which according to the base rate neglect bias, humans tends to ignore. Similar to the medical diagnosis problem, the base rate of lions is given as 1:1000 and therefore, based of Bayes theorem, the true positive rate is about 2%.

The tomato's disease classification question is a ML analogous to the medical diagnosis problem phrased with frequencies (natural number) instead of probabilities (percentage) as proposed by Cosmides and Tooby [30]. In this case, a ML algorithm is trained to detect diseased tomatoes. Again, 5% of the healthy tomatoes are detected as diseased, but the question phrasing is that 5 out of 100 healthy tomatoes are detected as diseased. The base rate is also given as frequency, and 1:1000 is phrased as the disease attacks about 1 out of 1,000 bushes. The answer, that is the true positive rate, is the same as in the previous question, about 2%.

Table 1: The lion classification question

| |
|---|
| A machine learning algorithm was trained to detect photos of lions. The algorithm does not err when detecting photos of lions, but 5% of photos of other animals (in which a lion does not appear) are detected as a photo of a lion. The algorithm was executed on a dataset with a lion-photo rate of 1:1000. If a photo was detected as a lion, what is the probability that it is indeed a photo of a lion?<br>(a) About 2%<br>(b) About 5%<br>(c) About 30%<br>(d) About 50%<br>(e) About 80%<br>(f) About 95%<br>(g) Not enough data is provided to answer the question |

Table 2: The tomato's disease classification question

| |
|---|
| A machine learning algorithm was trained to detect a leaf disease in photos of tomato bushes. The algorithm detects the diseased bushes perfectly, but 5 out of 100 healthy bushes are also detected as diseased. The disease affects about 1 out of 1,000 bushes. If a bush is detected as diseased, what is the probability that it is really diseased?<br>    (a) About 2%<br>    (b) About 5%<br>    (c) About 30%<br>    (d) About 50%<br>    (e) About 80%<br>    (f) About 95%<br>    (g) Not enough data is provided to answer the question |

The research was conducted in two phases: In the first phase, to verify the existence of the base rate neglect bias in the context of human interoperability of ML algorithms, we only asked the lion question (see Table 1). In the second phase, we studied the effect of frequencies (instead of probabilities) phrased question on the prevalence of base rate neglect bias. Therefore, in the second phase we asked both the lion classification question and the tomato's disease classification question.

### 3.4 Research Population

The questionnaire was distributed to learners from a variety of backgrounds who were taking Introduction to ML courses. (Note: The course was not the same for all populations; each population was studying a designated course designed for it.) In the first phase, total of 98 learners answered our first questionnaire and in the second phase, a total of 174 learners answered our second questionnaire. In total, 272 students answered our questionnaires (See Table 3).

Table 3: Research population

| | Number of participants | | |
|---|---|---|---|
| Population | Phase 1 | Phase 2 | Total |
| Undergraduate computer science electrical engineering and data science students | 51 | 68 | 119 |
| Computer science high school teachers | 26 | 34 | 60 |
| Social sciences and digital humanities researchers | 21 | 51 | 71 |
| **Total** | **98** | **153** | **251** |

## 4 RESULTS

Figure 3 presents the results of lion's classification question in the first stage. As can be seen, the ML learners exhibited the same patterns found by Casscells et al. [26]: the majority ignored the base rate and answered the lion classification question incorrectly (61%); and, among them, the most common erroneous response was 95% (given by 32% of the participants in the pilot phase). While Casscells et al. [26] found that 18% of the participants gave the correct answer, in our experiment, 39% answered this question correctly, that is, more than double than the rate found by Casscells et al. [26]. This difference can be explained in two ways. First, the question was asked as a multiple-choice question (rather than open one as in Casscells et al's case), providing the participants with several options. Thus, participants could consider the different answers or even guess the correct answer. The second explanation is derived from the participants' answers to the open question asked just after the lion's classification question, in which they were asked to explain their selected answer. More than half of the participants who answered the question correctly explained their answer by Bayes' rule by which they could calculate the answer precisely. It is, therefore, reasonable to assume that our population had more background in probability than did the populations of the Casscells et al.'s experiments.

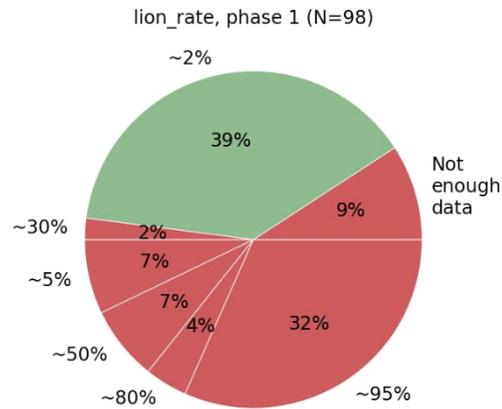

Figure 3: Lion Classification (phase 1): Distribution of Answers (n=98).

Following Cosmides and Tooby [30], in the second phase, we asked the two questions: the original lion's classification question and the tomato disease question, which is an identical question but phrased with frequencies instead of probabilities, i.e., natural numbers instead of percentage. The answers to these questions exhibited the same pattern found by Cosmides and Tooby [30] (see Figure 4). While in the original study [30] 56% answered the frequencies question correctly, in our research 58% answered this question correctly.

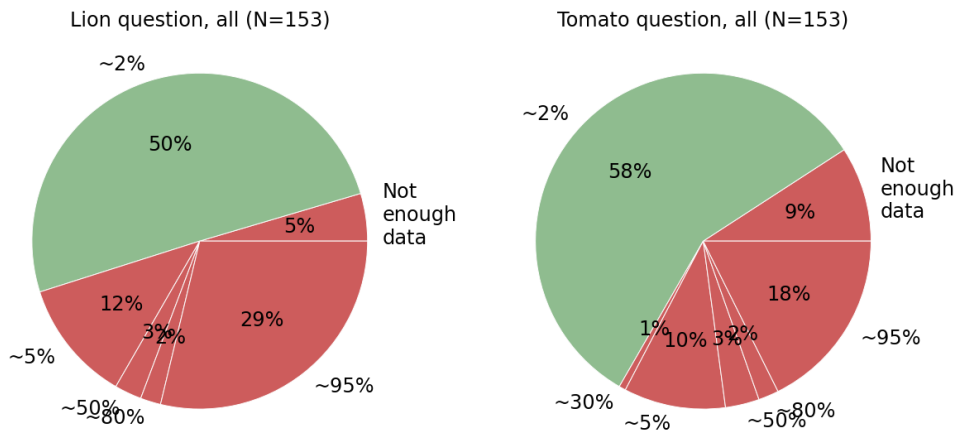

Figure 4: Lion Classification (left) vs. Tomato Disease Classification (right): Distribution of answers (n=153)

We also note that the participants in the second phase, who had both questions, exhibited a higher rate of success answering the lion's classification (probabilities) question (50%; see Figure 4) compared with participants who answered the first phase questionnaire in which only the probabilities (with percentages) formulation appeared (39%; see Figure 3). An explanation to this phenomenon can be find at Ejersbo and Leron [31], who suggest that the question phrased with frequencies (the tomato classification question in our case) bridged the intuitive and analytical thinking and helped participants reach a correct analytical solution for the probabilities question (the lion photos classification question in our case). We also suggest that since more students answered the question phrased with probabilities correctly when it was also phrased with frequencies, questions phrased with frequencies should be considered as part of ML courses.

Figure 5 presents the answer distribution of our three research populations: computer science, electrical engineering, and data science undergraduate students (upper row), social science and digital humanities researchers (middle row), and computer science high school teachers (lower row). For each population, the right column presents the answers to the lion classification question in Phase 1, the middle column presents the answers to the lion classification question in Phase 2 and the left column presents the answers to the tomato's disease classification question presented only in Phase 2. To simplify the presentation of the different distributions, instead of presenting the prevalence of each answer, all the wrong answers are unified and are represented by an "incorrect" category, and the 2% correct answer is represented by a "correct" category.

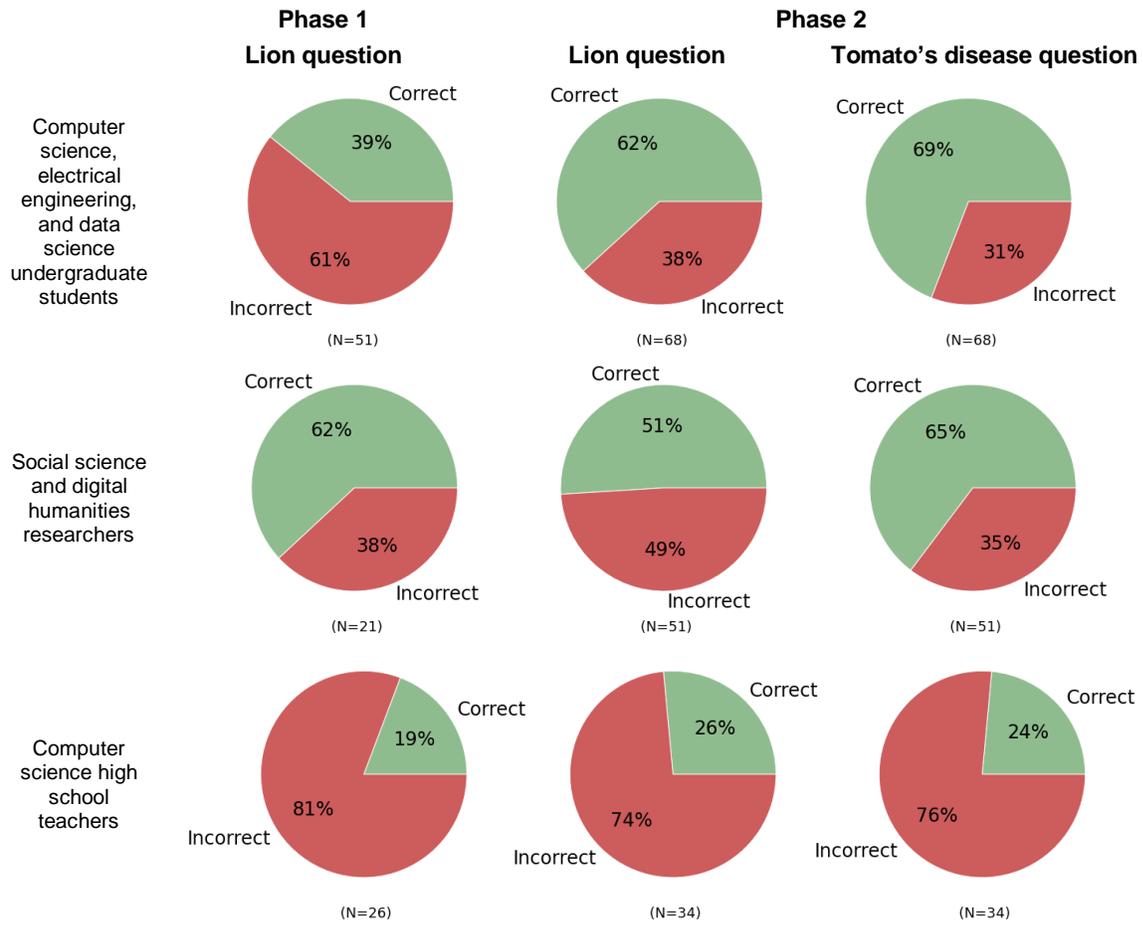

Figure 5: Lion Classification vs. Tomato Disease Classification: Distribution by Phase and Population

To compare the success rates between the populations and between the questions, we applied a statistical test to examine the differences between the distributions [32]. In our case, we were only interested in the success rate, and had no need to distinguish between the different types of errors. We therefore treated the different types of errors as one type of answers and designated all answers as one of two types: correct (~2%) or incorrect (all other answers). In this case, we could use the two-tailed binomial test to test the hypothesis that the distribution of the answers to the two questions is significantly different for the different populations, for each phase [33], [34].

Table 4 presents the distribution of answers by question and by population and the significance of the difference between the answers for each population (the lion question in both phases, and the line question vs. the tomato question in Phase 2). As can be seen:

- For the undergraduate students, the difference between the answers to the lion classification question in Phase 1 and Phase 2 is significant, while the difference between the answers to the lion classification question in Phase 2 and the answers to the tomato's disease question is not significant. This implies that for this population, the tomato's disease question indeed bridged the base neglect bias.
- For the social science and digital humanities researchers, the difference is not significant, as most of these participants are already familiar with Bayes law.
- For the computer science high school teachers, the difference is not significant as only a low percentage of this population solved the tomato's disease question correctly, and therefore bridging the base neglect bias could not be supported.

Table 4: Comparison of answer distribution by type of question

|  | Computer science, electrical engineering, and data science undergraduate students | Social science and digital humanities researchers | Computer science high school teachers |
|---|---|---|---|
| **Lion classification Phase 1 vs. Phase 2** | | | |
| Lion classification phase 1 | Correct:20 Incorrect: 31 | Correct:13 Incorrect: 8 | Correct:5 Incorrect: 21 |
| Lion classification phase 2 | Correct:42 Incorrect: 26 | Correct:26 Incorrect: 25 | Correct:9 Incorrect: 25 |
| Two tailed binomial value | p=0.00025 (*) | p=0.26 | p=0.27 |
| **Lion classification Phase 2 vs. tomato disease classification** | | | |
| Lion classification phase 2 | Correct:42 Incorrect: 26 | Correct:26 Incorrect: 25 | Correct:9 Incorrect: 25 |
| Tomato disease classification | Correct:47 Incorrect: 21 | Correct:33 Incorrect: 18 | Correct:8 Incorrect: 26 |
| Two tailed binomial value | p=0.26 | p=0.051 | p=0.84 |

(*) significant p<0.05

Table 5 presents the comparison between the distribution of the answers of the different populations. As can be seen:
- The difference between the answers of the computer science, electrical engineering, and data science undergraduate students and the answers of the social science and researchers is significantly different only with respect to their answers to the lion classification question in Phase 1. This again indicates that the undergraduate students, after solving the tomato's disease classification question in Phase2, were able to solve the lion question with a performance similar to that of the social science and digital humanities researchers.
- The computer science high school teachers population is significantly different from both computer science, electrical engineering, and data science undergraduate students and social science and digital humanities researchers in the three comparisons due to their low percentages of correct answers to the three questions.

Table 5: Comparison of answer distribution by population

| **Population 1** | **Computer science, electrical engineering, and data science undergraduate students** | **Social science and digital humanities researchers** | **Computer science high school teachers** |
| --- | --- | --- | --- |
| **Population 2** | **Social science and digital humanities researchers** | **Computer science high school teachers** | **Computer science, electrical engineering, and data science undergraduate students** |
| Lion classification phase 1 | p=0.04 (*) | p=0 (*) | p=0.04 (*) |
| Lion classification phase 2 | p=0.11 | p=0 (*) | p=0.005 (*) |
| Tomato disease classification | p=0.54 | p=0 (*) | p=0 (*) |

(*) significant p<0.05

## 5 DISCUSSION AND CONCLUSION

The ubiquity of ML algorithms in decision-making processes and the importance of ML interpretability and explainability in this context, prompt ML researchers to develop new methods to improve the interpretation and measurement of ML algorithm outcomes. ML interpretability and explainability have, however, two sides: machine and humans. Therefore, advancing ML interpretability and explainability is not only the responsibility of ML researchers, but also of ML educators.

In this paper, we showed that even though the percentage of correct answers in the first phase of our experiment is more than double that of the original experiment, the majority of the participants, who were at the time taking a ML course, did not answer the question correctly. This phenomenon is meaningful and alarming since the evaluation of a detector's prediction is a crucial stage in the data science workflow, based on which decision may be taken on its use in real life situations. Think about terrorism detection, different medical treatments, etc.

From this perspective, we mention that although in this paper we focused on empirical evidence of the misinterpretation of ML algorithms performance due to base rate neglect cognitive bias, we use the questions presented in the paper (Tables 1 and 2) not only for research but also for teaching. The questionnaire is used as a pre-lesson activity prior to the lesson on ML algorithms performance measurement. This pre-lesson activity reflects to the students of their own biases and thus may promote their understanding the broad considerations required in implementing ML algorithms is real-world applications. In addition, this activity leads to the introduction of the concept of confusion matrix (see Figures 1 and 2). Our results show that presenting the performance of ML algorithm using frequencies, that is natural numbers, for example in a confusion matrix, has the potential to mitigate the base rate neglect for computer science, electrical engineering, and data science undergraduate students. Special attention, however, should be given to Bayes theorem in teacher preparation programs for ML teaching. We call ML researchers and educators to develop and share more teaching methods to support teaching of this topic.

## ACKNOWLEDGMENTS

This work was supported by generous funding from the Technion's Machine Learning & Intelligent Systems (MLIS) Center.